\documentclass{PoS}
\usepackage{amsmath}

\newcommand{\e}{\mathrm{e}}

\title{The Planck and LHC results and particle physics}

\ShortTitle{The Planck and LHC results and particle physics}

\author{\speaker{Fedor BEZRUKOV}%
  \\
  Physics Department, University of Connecticut,\\
  2152 Hillside Road, Storrs, CT 06269-3046, U.S.A.\\
  RIKEN-BNL Research Center, Brookhaven National Laboratory,\\
  Bldg.\ 510a, Upton, NY 11973-5000, U.S.A.\\
  E-mail: \email{Fedor.Bezrukov@uconn.edu}}

\abstract{I will discuss the recent LHC and Planck results, which are
  completely compatible with the Standard Model of particle physics,
  and the standard cosmological model ($\Lambda$CDM), respectively.
  It turns out that the extension of the Standard Model is, of course,
  required, but can be very minimal.  I will discuss also what future
  measurements may be important to test this approach.}

\FullConference{The European Physical Society Conference on High Energy Physics \\
                18-24 July, 2013\\
                Stockholm, Sweden}

\begin{document}

\section{Introduction and overview of current experimental results}

Current results from LHC and Planck show incredible agreement with the
Standard Model of particle physics (SM) on the one hand and standard
cosmological model ($\Lambda$CDM) on the other hand.  Moreover, in
cosmology the Planck results \cite{Ade:2013ktc} constrain the
primordial perturbations to have low amplitude of tensor modes and
have no significant non-Gaussianities, thus favoring single field
inflation with relatively low inflationary scale.  In particle physics
a Higgs boson with the mass around 126\,GeV was discovered
\cite{CMS:2012gu,ATLAS:2012gk}, and further searches reveal no new
particle states and no deviations from the SM in predictions of rare
decays.

At the same time extension of the SM is necessary to explain
experimental problems, that are neutrino oscillations, baryon
asymmetry of the Universe, Dark Matter, and
inflation.\footnote{Strictly speaking, one should explain the observed
  nearly scale invariant spectrum of primordial perturbations.}  Usual
extensions of the SM introduce some particle physics at scales far
above electroweak scale.  In such quantum field theories large
(proportional to the mass of heavy particles) radiative contributions
to the mass of the Higgs boson emerge, which is one way to see the
hierarchy problem.  These contributions may be cancelled in
e.g.~supersymmetric theories, but until now no evidence of
supersymmetric partners for SM particles (or other deviations form SM)
have been observed.

An alternative solution to the hierarchy problem may be searched in
models without new particle physics above electroweak scale.  In this
case the only heavy scale in the theory is related to the Planck scale
defining the quantum gravitational effects, where the problem of
quadratic divergences may be different from the ordinary field theory
(see \cite{Dubovsky:2013ira} for a toy model example).  To attempt
this approach all the experimental facts mentioned in the previous
paragraph should be explained in a theory without heavy particles, see
talk by M.~Shaposhnikov \cite{Shaposhnikov:2013ira}.  The solution to
the Dark Matter and baryogenesis can be achieved within the $\nu$MSM
extension of the SM by three light sterile neutrinos, see talks by
M.~Shaposhnikov and O.~Ruchayskiy, and ref.~\cite{Boyarsky:2009ix} for
review.

This talk gives a short overview of selected simple inflationary
models that can be used in such a framework.  One way is to add a new
(light) scalar inflaton and make it consistent with observations by a
small non-minimal coupling, which is a rather conservative particle
physics model of inflation.  Second way is to use large non-minimal
coupling for the Higgs boson itself with the advantage of having no
additional degrees of freedom, but with potentially complicated
quantum dynamics at high energies.  The third method is the $R^2$
inflation, which provides the solution completely within the
gravitational part of the theory.

\section{Simple inflationary models}

\subsection{Light non-minimally coupled inflaton with quartic potential}

Probably the simplest and most widely known inflationary model uses an
additional dedicated scalar field \cite{Linde:1981mu}.  The proper
normalization of the density perturbations requires the potential of
the field to be very flat.  In case of the quartic potential this
defines the self coupling constant $\beta\sim1.5\times10^{-13}$.
However, the energy scale during inflation turns out to be not small
enough and significant amount of tensor perturbations is generated.
The recent Planck results exclude this model at more than 95\%
confidence level.  This does not mean that models with simple quartic
potentials should be abandoned as inflationary theories.  For a scalar
field an additional non-minimal coupling to gravity can be added to
the model \cite{Salopek:1988qh,Kaiser:1994vs}.  Let us note that such
a coupling is even required by the renormalization of a scalar field
in a curved space-time background.

The part of the action responsible for inflation is then
\begin{equation*}
  S_{\mathrm{NM}X^4} = \int\sqrt{-g}d^4x\left(
    - \frac{M_P^2+\xi X^2}{2}R
    + \frac{1}{2}\partial_\mu X\partial^\mu X
    - \frac{\beta}{4}X^4
  \right),
\end{equation*}
where we neglected the possibility of $X$ having a nonzero vacuum
expectation value contributing significantly to the Planck mass (see
\cite{Bezrukov:2013fka} for further discussion).  The simplest way to
analyze this model is to make the conformal transformation of the
metric
\begin{equation*}
  g_{\mu\nu} \to  \tilde{g}_{\mu\nu} = \Omega^2 g_{\mu\nu},
  \qquad
  \Omega^2 = 1+\xi X^2/M_P^2,
\end{equation*}
and redefine the inflaton field $X\to\chi$ to regain canonically
normalized kinetic term.  In the new variables $\tilde{g}_{\mu\nu}$
and $\chi$ (Einstein frame, as opposed to the original Jordan frame)
gravity couples minimally, while the inflaton potential gets rescaled
according to
\begin{equation}
  \label{eq:U(cX)}
  U(\chi) = \frac{\beta X^4(\chi)}{4\Omega^4(\chi)},
  \qquad
  \frac{d\chi}{dX} = \sqrt{\frac{\Omega^2+6\xi^2X^2/M_P^2}{\Omega^4}}.
\end{equation}
This potential is evidently more flat and gives less tensor
perturbations.

Starting from this point the analysis is a straightforward slow roll
inflation with one field.  It turns out that a very well analytic
approximation can be achieved by using the following relation for the
field value $X_N$ at the $N$ e-foldings \( (1+6\xi)X_N^2/M_P^2 =
8(N+1) \).  Though it is formally correct only at small $\xi$, it
smoothly interpolates between the small $\xi$ and large $\xi$ regimes
(for large $N$) \cite{Bezrukov:2013fca}.  The resulting slow roll
parameters can be obtained now by usual means
\cite{Bezrukov:2013fca,Salopek:1988qh,Kaiser:1994vs}, and interpolate
between zero $\xi$ and large $\xi$ limits (see fig.~\ref{fig:nsr}).
In particular, the tensor-to-scalar ratio $r$ drops rapidly with
increasing $\xi$ and becomes compatible with observations for
$\xi\gtrsim0.001$.  The self coupling constant $\beta$ is determined
by the CMB normalization and interpolates between the zero $\xi$ value
and the large $\xi$ case $\beta\simeq(47000/\xi)^2$.

The number of e-foldings is not in fact arbitrary for quartic
inflation, because after the end of the slow roll the evolution of the
Universe is dominated by the oscillations of the field in quartic
potential which corresponds to the radiation dominated expansion.
Thus, though the real reheating (transfer of the energy to the SM
degrees of freedom) is not be immediate, for the purposes of the
estimate of $N$ for the pivot scale $k/a=0.002$\,Mpc$^{-1}$ the
``reheating'' is immediate, leading to $N\simeq60$, see
\cite{Bezrukov:2013fca}.

Note that non-minimal coupling lessens the concerns that may appear in
the large field inflation because of the field spanning superplanckian
range during inflation.  For the minimally coupled inflation the field
changes by about $X_N\sim22M_P$, while for $\xi\sim1$ only by $8M_P$.
For $\xi>1$ the situation becomes more complicated because of the
additional scales $M_P/\xi$ and $M_P/\sqrt{\xi}$ appearing in the
model.

To complete the model one should couple the inflationary sector to the
SM.  Here we would like to make an additional assumption that the
scale invariance is only broken in the inflaton sector.  Then the
scalar part of the model Lagrangian density becomes
\begin{equation}
  \label{XHint}
  \mathcal{L} =
  \frac{1}{2}m_X^2X^2 - \frac{\beta}{4}X^4
  -\lambda\left(H^\dagger H-\frac{\alpha}{\lambda}X^2\right)^2,
\end{equation}
where $H$ is the Higgs field doublet and $\alpha$ is the only constant
controlling the interaction between the SM and inflationary sector.
The only free parameter here is $\alpha$ ($\lambda$ is determined by
the Higgs boson mass, and $\beta$ by the CMB normalization).  From the
requirement of good cosmological evolution $\alpha$ can not be too
large not to spoil the inflationary potential with radiative
corrections; it can not be too small in order to get sufficient
reheating temperature (here we mean transfer of energy into the SM
degrees of freedom) to allow for transfer of the asymmetry from the
lepton to the baryon sector by the sphaleron transitions
\cite{Burnier:2005hp}.  The resulting bounds on $\alpha$ are
\(
  0.7\times10^{-11} \lesssim \alpha \lesssim \sqrt{0.1\times\beta}
\),
see \cite{Anisimov:2008qs}.  For the potential (\ref{XHint}) these
range translates into the bounds on the mass of the $X$ particle
\cite{Bezrukov:2009yw,Bezrukov:2013fca}.  The resulting inflaton has
the mass around GeV and can be searched in rare decay experiments on
LHCb.

Another consequence of (\ref{XHint}) is that the inflation proceeds
along the line $H^\dagger H=\frac{\alpha}{\lambda}X^2$ in the field
space.  Thus, the value of the Higgs field $H$ is actually rather
large at inflation, of the order $\sim\sqrt{\alpha/\lambda}M_P$.  This
implies the lower bound on the Higgs boson mass similar (but just
slightly weaker) than that of the requirement of the absolute
stability of the electroweak vacuum (or, equivalently, viability of
the Higgs inflation)
\cite{Bezrukov:2012sa,Degrassi:2012ry,Alekhin:2012py}.

\subsection{Higgs inflation}

For a large value of $\xi$ the the self-coupling constant can be made
arbitrarily large.  So, instead of adding a new field to the model it
is possible to just add the non-minimal coupling to gravity of the
Higgs field itself \cite{Bezrukov:2007ep}
\[
  S_{\mathrm{NMHiggs}} =
  \int \sqrt{-g} d^4x\left(-\xi_h H^\dagger HR\right).
\]
The HI case corresponds to the large $\xi_h\gg1$ and negligible vacuum
contribution of the non-minimal coupling to the Planck mass, $\xi_h
v^2\ll M_P^2$.  In the unitary gauge $H=\frac{1}{\sqrt{2}}{0 \choose
  v+h}$ the inflationary analysis is analogous to the previous section
with the obvious substitution $X\to h$ and $\beta\to\lambda$.  For the
large Higgs background $h,\chi\gg M_P/\xi_h$ (relevant for inflation)
the potential is
\begin{equation}
  \label{U(chi)}
  U(\chi) \simeq \frac{\lambda M_P^4}{4\xi_h^2} \left(
    1 - \e^{-\frac{2\chi}{\sqrt{6}M_P}}
  \right)^2 .
\end{equation}
For small field $h,\chi\ll M_P/\xi_h$, the model returns to the usual
SM regime.

\begin{figure}
  \centerline{%
    \raisebox{-0.003\textwidth}{\includegraphics[width=0.364\textwidth]{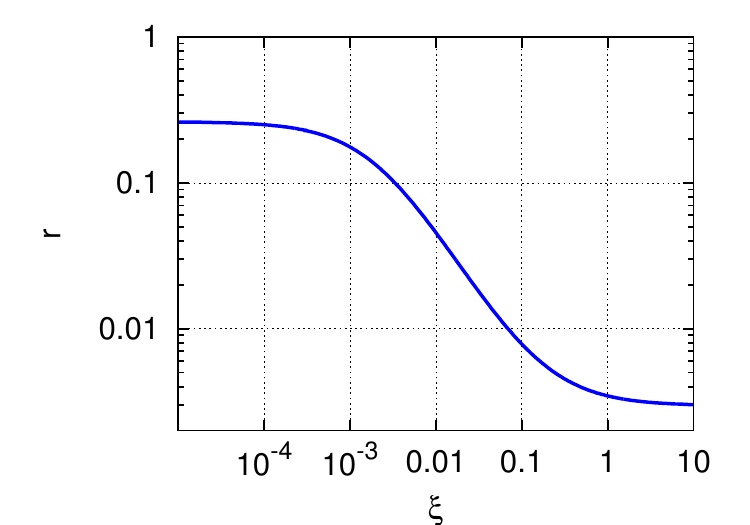}}
    \includegraphics[width=0.5\textwidth]{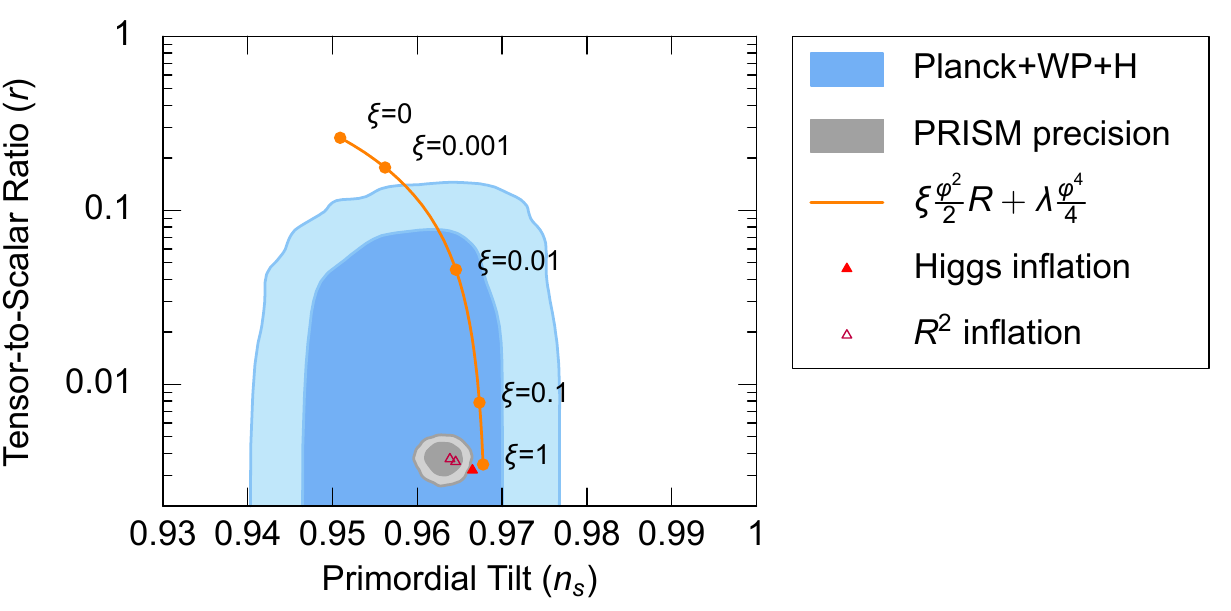}}
  \caption{Left: dependence of the tensor-to-scalar ration $r$ on
    non-minimal coupling constant $\xi$ for the quartic potential.
    Right: predictions of the described inflationary models for
    tensor-to-scalar ratio $r$ and spectral index $n_s$ and existing
    experimental bounds from Planck 2012 release \cite{Ade:2013ktc}.
    Also shown expected precision from the PRISM experiment
    \cite{Andre:2013afa}.}
  \label{fig:nsr}
\end{figure}

The CMB normalization requirement fixes the non-minimal coupling
\(
  \xi_h \simeq 47000\sqrt{\lambda}
\),
where $\lambda$ is the Higgs boson self-coupling constant taken at the
inflationary scale.  To the lowest order in $1/\xi$, the spectral
index and the tensor-to-scalar perturbation ratio are
\(
  n_s \simeq 1-8{(4N+9)}/{(4N+3)^2} \simeq 0.967
\),
\(
  r \simeq {192}/{(4N+3)^2} \simeq 0.0031
\).
The inflation model in the Einstein frame is
a simple one field slow-rolling inflation, with all extra degrees of
freedom much heavier than the Hubble scale ($m\sim M_P/\sqrt{\xi_h}\gg
H\sim M_P/\xi_h$), so it does not predict any significant
non-Gaussianities in the spectrum, and the values of the parameters
are well in agreement with observations, see fig.~\ref{fig:nsr}.

Note that reheating is quite effective in this model
\cite{Bezrukov:2008ut,GarciaBellido:2008ab}, with
$T_r\sim0.3\text{--}1.1\times10^{14}$\,GeV.  Precise estimate is
complicated, but in any case for $h<M_P/\xi$ the expansion of the
Universe is governed by the quartic potential of SM and is radiatively
dominated, so effectively $T_r\gtrsim10^{13}$\,GeV
\cite{Bezrukov:2013fka}.  This reheating temperature corresponds to
the number of e-foldings $N\simeq 57.5\pm0.2$.

The calculation of the quantum corrections in the Higgs inflation
requires additional assumptions on the UV properties of the theory.
The variable change (\ref{eq:U(cX)}) leads to higher dimensional
operators in the potential suppressed by the scale $M_P/\xi$ leading
to tree unitarity violation in electroweak vacuum at this energy.
This scale is just slightly above to the Hubble scale at inflation
$H\sim\sqrt{\lambda}M_P/\xi$.  The situation is improved by noting
that at inflation the background is very different from the vacuum
solution and the small perturbations on top of this background can be
safely analyzed at tree level up to a much higher energy (in the
Einstein frame up to $M_P$ for the gravitational perturbations, and
$M_P/\sqrt{\xi}$ for SM particle like excitations, see
\cite{Bezrukov:2013fka,Bezrukov:2010jz}).  The inflation starts from
the large value of the Higgs field the Higgs inflation can proceed
only if the electroweak vacuum is absolutely stable, providing a lower
bound on the Higgs mass \cite{Bezrukov:2012sa}.

\subsection{$R^2$ inflation}

A related mechanism of inflation emerges in modification of the
gravity action alone \cite{Starobinsky:1980te,Gorbunov:2010bn}
\begin{equation*}
  S_{R^2} =
  \int\sqrt{-g}d^4x \left(
    - \frac{M_P^2}{2}R + \frac{M_P^2}{6\mu^2}R^2
  \right).
\end{equation*}
As far as this is a theory with higher order derivatives it has an
additional degree of freedom, scalaron $\phi(x)$, which emerges in the
simplest way in the conformally transformed (Einstein) frame
\begin{equation*}
  g_{\mu\nu}\to\tilde{g}_{\mu\nu}=\e^{\sqrt{2/3}\phi/M_p}g_{\mu\nu},
\end{equation*}
with the Einstein frame action having \emph{exactly} the form
\begin{equation}
  \label{eq:5}
  S_{EF} = \int\sqrt{-\tilde{g}}d^4x\left\{
    -\frac{M_P^2}{2}\tilde{R}+\frac{1}{2}\partial_\mu\phi\partial^\mu\phi
    -\frac{3\mu^2 M_P^2}{4}\left(1-\e^{-\frac{2\phi}{\sqrt{6}M_P}}\right)^2
  \right\}.
\end{equation}
In the inflationary domain this potential coincides with that of the
Higgs inflation (\ref{U(chi)}), so the inflationary predictions of
these models are very similar.  However, the reheating process here
proceeds via Planck scale suppressed operators
\cite{Gorbunov:2010bn,Gorbunov:2012ns} leading to significantly lower
reheating temperature $T_r\sim3.1\times10^9$\,GeV (or even lower for
conformally coupled Higgs boson \cite{Gorbunov:2012ns}).  This leads
to a different number of e-foldings $N\simeq54$
\cite{Bezrukov:2011gp,Gorbunov:2012ns} and slightly modified CMB
predictions, see fig.~\ref{fig:nsr}.  The next generation CMB
experiments may be able to reach the required precision.

Another important feature of the $R^2$ inflation is that it does not
require the Higgs field to reach large values throughout the evolution
of the Universe, thus making it possible to work even if the
electroweak vacuum is metastable
\cite{Bezrukov:2011gp,Gorbunov:2012ns}.  Also, because the potential
in (\ref{eq:5}) is exact the cut-off scale in $R^2$ inflation is just
$M_P$, what is immediately seen after expanding the potential in power
series.  Note, that this effect is similar to what happens in the
inflationary models with non-minimal coupling in the induced gravity
regime.  (in particular the perturbative UV-completion of the Higgs
inflation \cite{Giudice:2010ka}).

\section{Conclusions}

I described here three inflationary models which can be realized on
top of the SM without introduction of heavy particle states in the
theory.  This is important if one wants to evade quadratic divergences
associated with heavy particles in the theory, and may allow to
connect inflationary physics with low energy phenomenology.

The main way to distinguish these models is the improved measurements
of the parameters of the primordial perturbations, that is the
spectral index and tensor-to-scalar ratio.  The not so small
tensor-to-scalar ratio can easily distinguish the non-minimally
coupled quartic inflation, and possibly even determine the value of
the non-minimal coupling $\xi$.  Significant improvement in $n_s$
measurement may distinguish $R^2$ and Higgs inflation.

Another significant measurement is related to the analysis of the
stability of the electroweak vacuum.  Specifically, the Higgs
inflation and light non-minimally coupled inflation described in this
talk are possible only if the electroweak vacuum is absolutely stable,
while the $R^2$ inflation may work even for a metastable one.  The
current Higgs boson mass can be compatible within experimental and
theoretical errors with both stable and metastable electroweak
vacuum \cite{Bezrukov:2012sa,Alekhin:2012ig}.  The most significant
improvement here may be measurement of the top quark mass (or directly
the top quark Yukawa constant), strong gauge coupling and, finally,
Higgs boson mass itself.

\bibliographystyle{JHEP}
\bibliography{Papers}

\end{document}